# Characterization of Monoenergetic Neutron Reference Fields with a High Resolution Diamond Detector

*A. Zimbal, L. Giacomelli, R. Nolte and H. Schuhmacher*

*Physikalisch-Technische Bundesanstalt,
Bundesallee 100, 38116 Braunschweig, Germany*

A novel radiation detector based on an artificial single crystal diamond was used to characterize in detail the energy distribution of neutron reference fields at the Physikalisch-Technische Bundesanstalt (PTB) and their contamination with charged particles. The monoenergetic reference fields at PTB in the neutron energy range from 1.5 MeV up to 19 MeV are generated by proton and deuteron beams impinging on solid and gas targets of tritium and deuterium. The energy of the incoming particles and the variation of the angle under which the measurement is performed produce monoenergetic reference fields with different mean energies and line shapes. Well established simulation codes allow these parameters to be calculated in detail, provided the properties of the targets are known.

In this paper we present high resolution neutron spectrometry measurements of different monoenergetic reference fields. The results are compared with calculated spectra taking into account the actual target parameters. The influence of deviations from the ideal case, e.g. a non homogeneous tritium distribution in a solid Ti/T-target, was investigated. Line structures in the order of 80 keV for a neutron energy of 9 MeV were resolved. The shift of the mean energy and the increasing of the width of the neutron peak with increasing pressure in the gas target in the order of 30 keV were measured.

Another result is the determination of the contamination of the neutron field at 14 MeV with high energy charged particles (protons) from side reactions inside the T-target. This effect is due to the thin backing of the targets in use at PTB. It depends on the age of the target and it has to be taken into consideration for irradiations at small distances for some detectors, especially when very old targets are used.

The experiments have shown that this detector is an easy to operate compact neutron spectrometer with extremely good energy resolution and that detailed structures in the line shapes of monoenergetic neutron fields can be resolved without using time of flight techniques.

**Dr. Andreas Zimbal
Physikalisch-Technische Bundesanstalt
6.5 - Neutron Radiation
Bundesallee 100
D-38116 Braunschweig, Germany
Phone: +49-531-592-6540
Fax : +49-531-592-6505**

**Andreas.Zimbal@ptb.de**

Introduction

In recent years, semiconductor detectors made from artificial chemical vapour deposition single crystal diamond (scCVD) have become commercially available. These detectors are very promising candidates for high resolution neutron spectrometry in very high neutron fluxes, which are expected e.g. in fusion research at the new ITER experiment, presently being built in Cadarache [1, 2]. Neutrons interact with diamond detectors through different (n,n´) and (n,$\alpha$)-reactions on Carbon. The $^{12}$C(n,$\alpha$)$^{9}$Be-reaction can be used to directly determine the line structure of monoenergetic neutron fields for neutrons with energies higher than 6 MeV. For a sufficiently large diamond crystal, the secondary charged particles are stopped in the crystal, and the total kinetic energy is deposited in the crystal. Therefore, this reaction produces a peak structure in the measured pulse height spectrum (PHS) which can be converted into the corresponding neutron energy spectrum just by a shift in the energy scale corresponding to the Q-value of the reaction of 5.702 MeV.

**Detector**

A scCVD detector from "Element Six Technologies", UK [3] (4 mm x 4 mm x 0.5 mm, impurities < 5 ppb N) was tested under different conditions with $\alpha$, electron and neutron radiation. The detector was fully contacted by the manufacturer without additional layers of converter material. No instructions for operation (such as high voltage (HV) and polarity) were given. The operational conditions had to be determined using sources prior to the neutron measurements. The charge collection properties of the crystal can best be determined using monoenergetic electrons (e.g. conversion electrons from a $^{207}$Bi-source) or $\alpha$-particles. Whereas the electrons deposit their energy throughout the whole volume of the crystal, $\alpha$-particles have a short path length of only 15 $\mu$m in the detector. The housing of the crystal is made in such a way that irradiations from the front and the back side, relative to the charge collection electrode, are possible. The irradiation from the back side with $\alpha$-particles is the most severe test of the charge collection properties of the crystal: The $\alpha$-particles are stopped immediately in the crystal and the charge carriers (electrons and holes) have to drift $\sim$ 500 $\mu$m before they reach the opposite contact. Stability issues can be investigated best using this irradiation condition. The homogeneous distribution of the charge during irradiation with monoenergetic electrons ($E_{max}$ < 1 MeV) is a good method for understanding the behaviour of the crystal under neutron irradiation, as the electrons deposit their energy in the whole crystal due to the low stopping power. Neutrons interact with the whole volume of the crystal, but their charged reaction products (recoil $^{12}$C, $\alpha$, $^{9}$Be) are stopped within a very short range.

**Charged particles measurements**

For checking the optimal performance of the crystal, standard analogue electronics consisting of a charge sensitive preamplifier (Ortec 142A) and a shaping main amplifier (Ortec 571, $\tau$ = 0.5 $\mu$s) together with a conventional data acquisition system (Analog-to-Digital converter + PC) was used. This limits the total count rate of the system but allows a direct comparison with a standard Si-diode. The results of this comparison, which was done similar to [4], are not discussed here in detail. The diamond crystal investigated permits the collection of electrons and holes (depending on the polarity of the collection voltage). Slightly higher signal amplitudes were found for the collection of electrons ($\sim$ 2.9 %, positive HV used) compared to negative HV. With increasing operational voltage the peak positions in the spectra raised and the width of the peaks decreased. The optimal operational voltage of the detector for both polarities was determined by successively increasing the HV until the peaks (for electron and $\alpha$-measurements) were found to be stable for at least some minutes to a few hours. For all measurements shown in this paper, a positive HV of 180 V was used. As examples, the PHS for measurements with a $^{207}$Bi-source and a triple $\alpha$-source ($^{239}$Pu, $^{241}$Am, $^{244}$Cm) are given in Fig. 1 and Fig. 2.

**Neutron measurements and applications**

For the investigation of the performance of the crystal for the measurement of neutron radiation, several irradiations with neutrons with energies from 2.0 MeV to 15 MeV were done. A demonstration of the capabilities of the detector as high resolution neutron spectrometer is given in Fig. 3, where the pulse height response for 14.0 MeV monoenergetic neutrons is shown. The peak due to the $^{12}C(n,\alpha)^9Be$-reaction is clearly visible at the right end of the spectrum. A precise understanding of all of the structures in the PHS needs further investigations and it will be the scope of future work aiming at a full characterization of the detector. If a response matrix can be determined for the whole neutron energy range from about 2 MeV to more than 14 MeV, the information in all channels of the PHS can be used to perform broadband neutron spectrometry applying unfolding methods. This was already demonstrated very succecfully for liquid organic scintillation detectors based on (n,p)-scattering at the JET facility for nuclear fusion research [5]. A more detailed view on the peak structure caused by 14 MeV neutrons is shown in Fig. 4. The PHS for measurements under different angles of the detector relative to the incoming deuteron beam of the accelerator show excellent agreement in the peak region with calculations performed with the PTB TARGET code [6]. The TARGET code calculates the neutron energy spectrum produced by deuterons impinging on a solid Ti(T)-target by considering the transport of charged particles (angle and energy straggling) in the target and the contribution of target scattered neutrons to the neutron energy spectrum at the position of the measurement. The energy scale of the PHS (Fig. 4) was calculated using a pulser calibration to determine the electronic offset and an adjustment to fit the 98° neutron measurement at 14.0 MeV. For this adjustment, a shift of the corresponding neutron energy scale (upper abscissa) of 5.702 MeV relative to the pulse height scale, caused by the Q-value of the $^{12}C(n,\alpha)^9Be$-reaction, was considered. No significant quenching was found for the different charged particles (e, $\alpha$, $^9Be$) within the limits of a linear energy scale given by the linearity of the components used for these experiments.

The detector is also well suited to determine a potential contamination of neutron fields by charged particles. For the PTB T-targets with thin backing (0.5 mm Al) there is the possibility that protons originating from the $^3He(d,p)^4He$-reaction ($E_p$ = 15 MeV) can escape from the target and hit a detector positioned at a close distance [7]. $^3He$ is accumulated in T-targets by β-decay of T ($t_{½}$ = 12.3 y). Therefore, the contamination with protons depends on the age of the target. This is demonstrated in Fig. 5. An old T-target (AL-95-2) shows a much higher contribution of protons (huge broad peak at about 10 MeV) than a new one (PTB06). To shield the detector from these protons, both measurements were repeated with a 1 mm thick Al-plate in front of the diamond crystal. In this case, the PHS for the old and the new target overlap (Fig. 6). The small insert of Fig. 6 shows the yield for different positions of the ion beam on the rotating targets. For the target AL-95-2 a huge inhomogeneity of more than a factor of 2 (compared to 20 % for the target PTB06) is clearly visible. Due to the low deuteron energies of about 200 keV used to produce the 14.8 MeV neutron field, time-of-flight methods using a pulsed beam could not be employed to study target properties. Therefore is was not clear up to now, if the neutron energy spectrum produced using the old inhomogeneous T-target is identical to the one produced with a more recent target. The direct comparison of the two targets of different age leads to the conclusion that although the target AL-95-2 is very inhomogeneous over the rotation circle of the d-beam of about 1 cm, the depth distribution of T inside the Ti-layer (0.957 mg/cm$^2$, ~ 2.1 μm) is homogeneous. All neutron measurements shown in Fig. 3 and 4 and those discussed below were made with a 1 mm Al-plate in front of the detector to avoid this influence.

Another application of this detector is the measurement of the line width of monoenergetic neutron fields produced by the $D(d,n)^3He$-reaction with a deuterium gas target (thickness = 30 mm). This reaction is used at the PTB accelerator facility to produce neutrons with energies from 5 to 15 MeV. For a deuteron energy of $E_d$ = 6.52 MeV, the resulting neutron energy is 9.4 MeV. By varying the pressure (515 hPa, 1019 hPa, 1545 hPa) in the gas target, the resulting width (FWHM) of the neutron energy spectrum was 82 keV, 92 keV and 112 keV (Fig 7). The calculated values for this target using the SINENA [8] code are 57 keV, 71 keV and 94 keV. This difference is due to the fact that the intrinsic resolution of the detector has not been considered yet.

The PHS spectra for irradiations with neutrons from 2.0 MeV to 2.8 MeV are given in Fig. 8. In this case the PHS is caused by C-recoils from elastic scattering. The details in the structure of the PHS reflect the structures of the scattering cross section at these energies. A resonance at $E_n$ = 2.08 MeV might be responsible for the more steeper PHS at 2.15 MeV neutron energy. The shapes of these PHS are very similar to those of liquid scintillation detectors irradiated with 2.5 MeV neutrons. This promises the possibility of applying unfolding methods as already been developed for these detectors [5].

An important parameter for the application of this detector as spectrometer in very high intensity neutron fields (e.g. at ITER) is its absolute efficiency. This value was determined from the known absolute neutron fluence values for the PTB reference neutron fields and the count rates measured in the PHS. Considering all events with an energy threshold of $E$ > 0.1 MeV, the response $R$ of the detector for 2.5 MeV and 14 MeV neutrons is comparable (see Tab. 1). For the 14 MeV measurements, the response in the peak region of the PHS (pulse height energy range from 8.1 MeV to 8.6 MeV, see Fig. 3 and 4) is nearly two orders of magnitude smaller (Tab. 1). Given the value of the expected fluence rate at the detector positions in the radial neutron camera of ITER ($\dot{\phi}$ = 3 · 10$^8$ cm$^{-2}$ s$^{-1}$) [9], the count rate for such a detector would be ∼ 6 · 10$^5$ s$^{-1}$ which can be handled by modern fast digital data acquisition systems based on direct signal sampling [10]. For such a setup however, the slow charge sensitive preamplifier has to be substituted with a low noise current amplifier with appropriate signal shaping. This is possible because of the negligible dark current and the fast rise time of the signals (a few ns) for this detector.

**Conclusion**

Artificial scCVD diamond detectors operated with standard analogue electronics can be used as high resolution semiconductor detectors. The charge collection of electrons and holes give comparable resolution in the PHS, but it was found that electron collection is more stable and allows a higher operational voltage. The response of commercially available detectors is stable for a few hours when irradiated with electrons (front and back side) or α-particles (front side, i.e. through the charge collection contact). For neutron measurements, certain points concerning the stability of the PHS need to be investigated further. Whereas for the 2.5 MeV neutron measurements the shape of the measured PHS remains stable, the efficiency decreases during operation with nominal high voltage after a few hours. For the 14 MeV neutron measurements it was observed that the ratio of peak/total-efficiency (Tab. 1) decreases during operation. First measurements indicate that these effects depend on the neutron fluence rate during the irradiation. All instability effects are completely reversible after switching off the HV. The peak structure visible for neutron energies > 8 MeV allows a direct characterization of neutron targets/beams, also in those cases where time-of-flight methods are not applicable. If a response matrix can be determined, broadband neutron spectrometry is possible via unfolding in the energy range from < 2 MeV to around 50 MeV. The accessible neutron energy range is limited at the low energy side by additional interactions of the detector with γ-radiation and on the high energy side by the range of α-particles for full charge deposition in the crystal.

| $E_n$ | $R_{peak}$ cm$^2$ | $R_{total}$ ($E$ > 100 keV$_{ee}$) cm$^2$ |
|---|---|---|
| 2.45 MeV | – | 2.1 · 10$^{-3}$ |
| 14.0 MeV | 4.7 · 10$^{-5}$ | 1.9 · 10$^{-3}$ |

Table 1: Fluence response $R$ of the scCVD diamond detector (4 mm x 4 mm x 0.5 mm) for neutrons of $E_n$ = 2.45 MeV and 14.0 MeV.

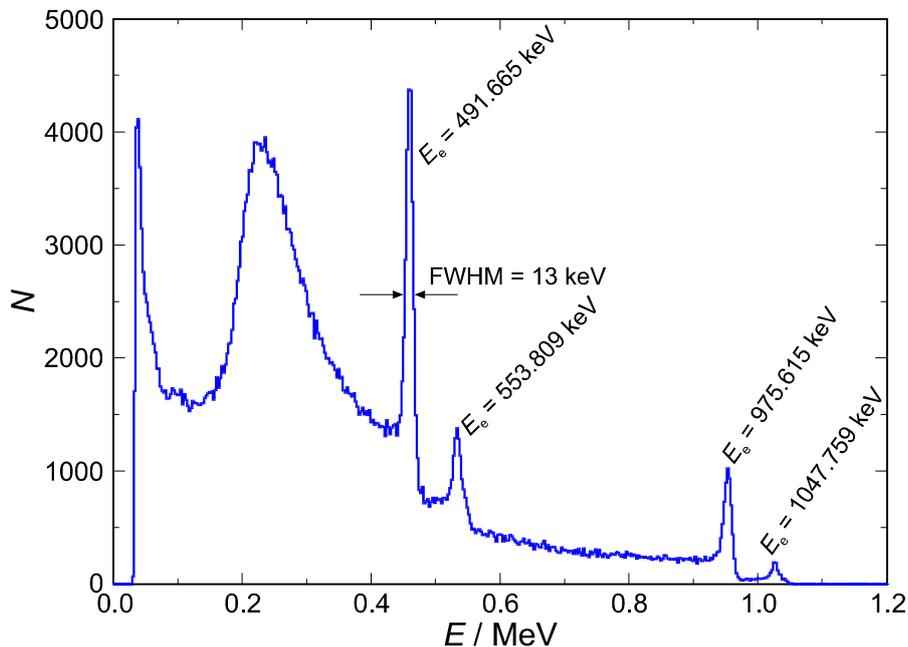

Figure 1: Number of counts, *N*, as a function of the energy, *E*, deposited in the detector for a $^{207}$Bi-conversion electron source (covered with 2.2 mg/cm$^2$ polyester), measured in air.

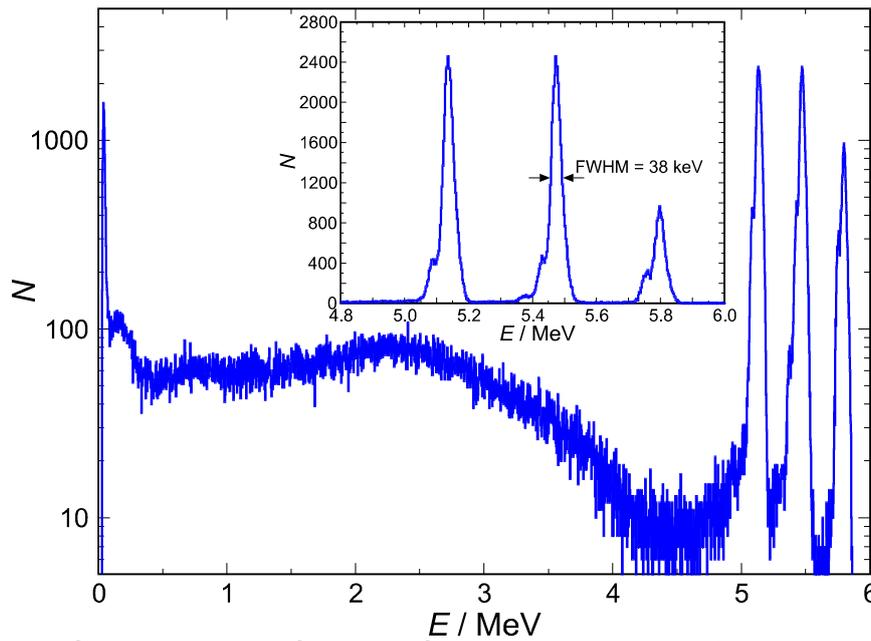

Figure 2: Number of counts, *N*, as a function of the energy, *E*, deposited in the detector for a mixed α-source ($^{239}$Pu, $^{241}$Am, $^{244}$Cm), measured in vacuum.

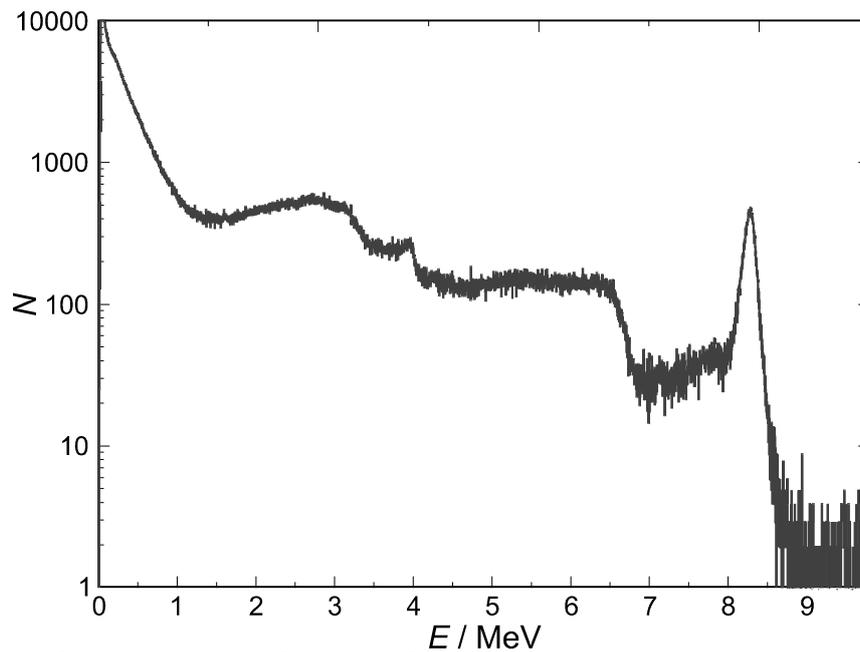

Figure 3: Number of counts, *N*, as a function of the energy, *E*, deposited in the detector measured for 14.0 MeV neutrons from the T(d,n)$^4$He-reaction at an angle of 98° relative to the incoming deuteron beam ($E_d$ = 215 keV).

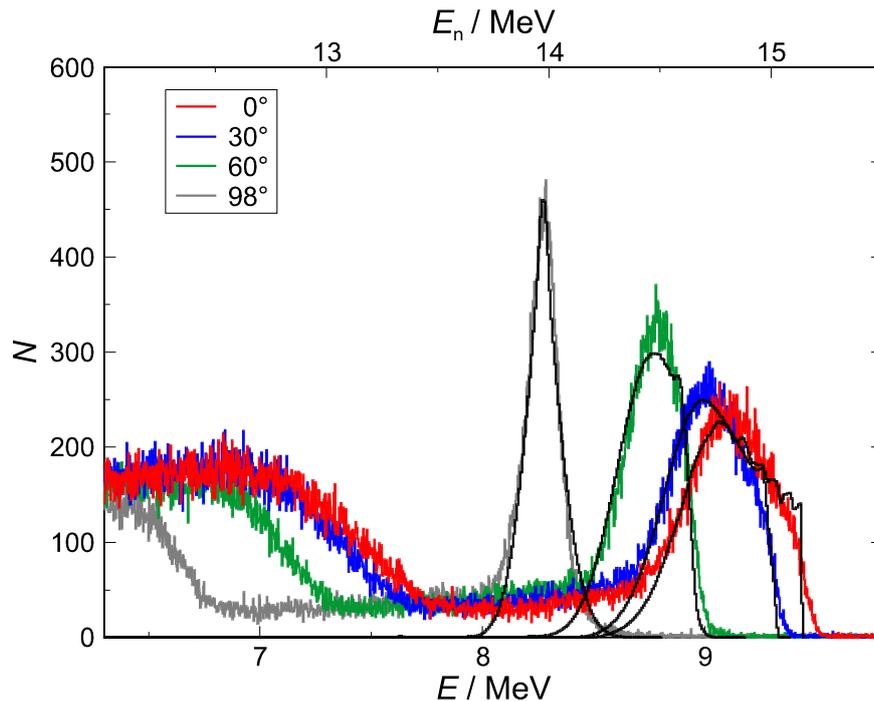

Figure 4: Number of counts, $N$, as a function of the energy, $E$, deposited in the detector measured for neutrons from the T(d,n)$^4$He-reaction for different angles relative to the incoming deuteron beam ($E_d$ = 215 keV). The measurements are compared to calculations with the TARGET code (black histogram, see upper abscissa). All measurements are normalized to the same number of counts from a monitor detector.

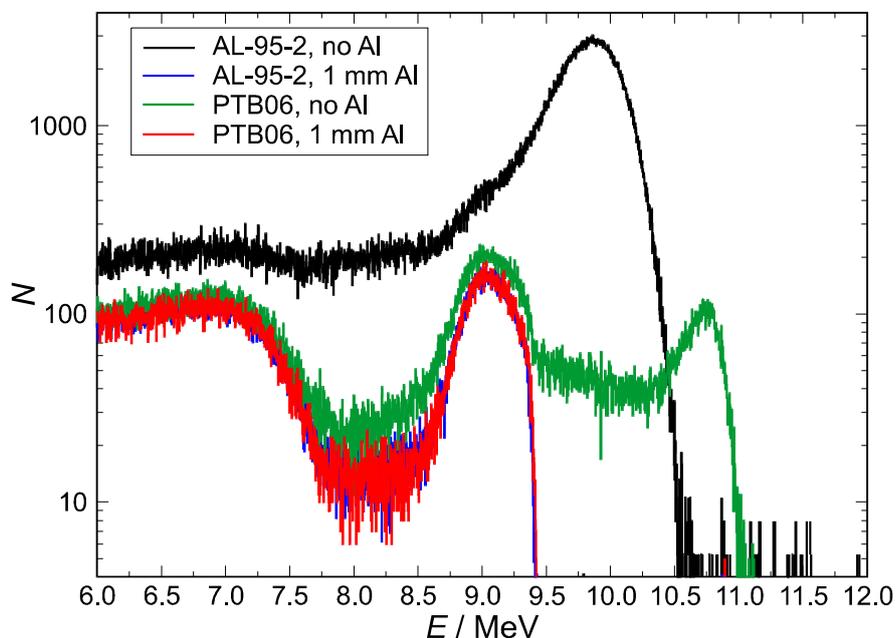

Figure 5: Number of counts, $N$, as a function of the energy, $E$, deposited in the detector measured for the T(d,n)$^4$He-reaction ($E_d$ = 215 keV) at a distance of 10 cm for two different Ti(T)-targets of different age. For a detector shielded with 1 mm Al, only neutron interactions are measured and the distributions agree. Without Al shielding, a contamination of the radiation field with high energy protons is clearly visible, in particular for the old target. All measurements are normalized to the same number of counts from a monitor detector.

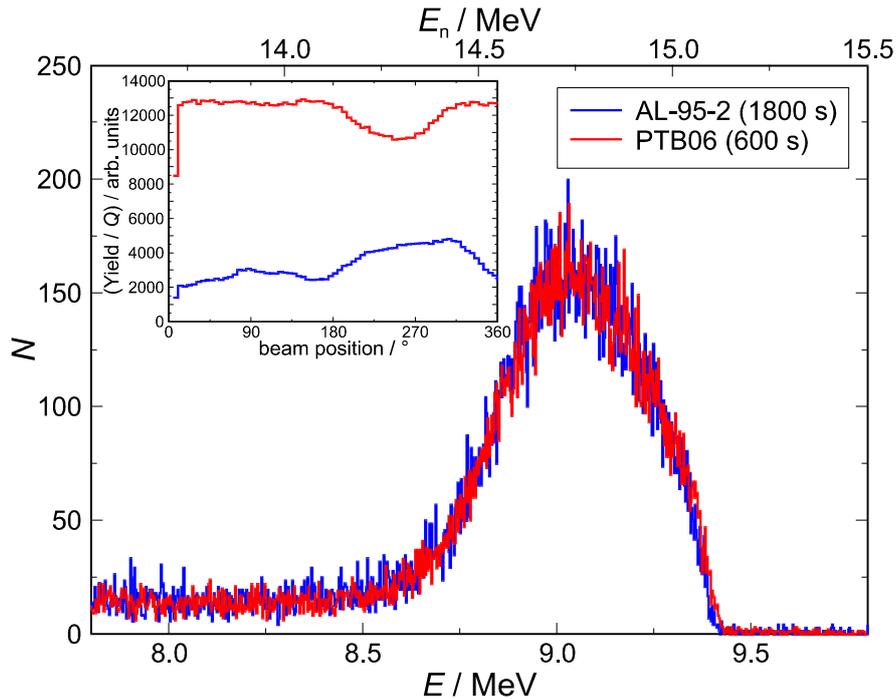

Figure 6: Number of counts, *N*, as a function of the energy, *E*, deposited in the detector measured for neutrons from the T(d,n)$^4$He-reaction ($E_d$ = 215 keV) for two Ti(T)-targets of different age. The old target (Al-95-2) shows huge inhomogeneities in the T-load compared to the new one (PTB06), see insert. Both measurements are normalized to the same number of counts from a monitor detector.

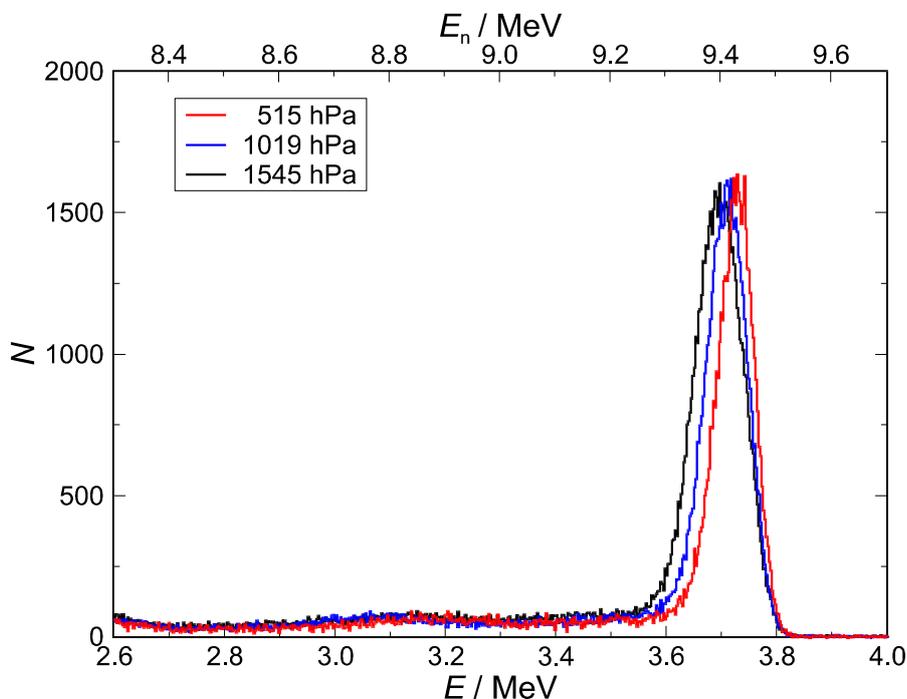

Figure 7: Number of counts, *N*, as a function of the energy, *E*, deposited in the detector for 9.4 MeV neutrons from the D(d,n)$^3$He-reaction in a deuterium gas target for different pressure and the same incoming deuteron energy ($E_d$ = 6.52 MeV). For better comparison, the measurements are normalized to the same peak amplitude.

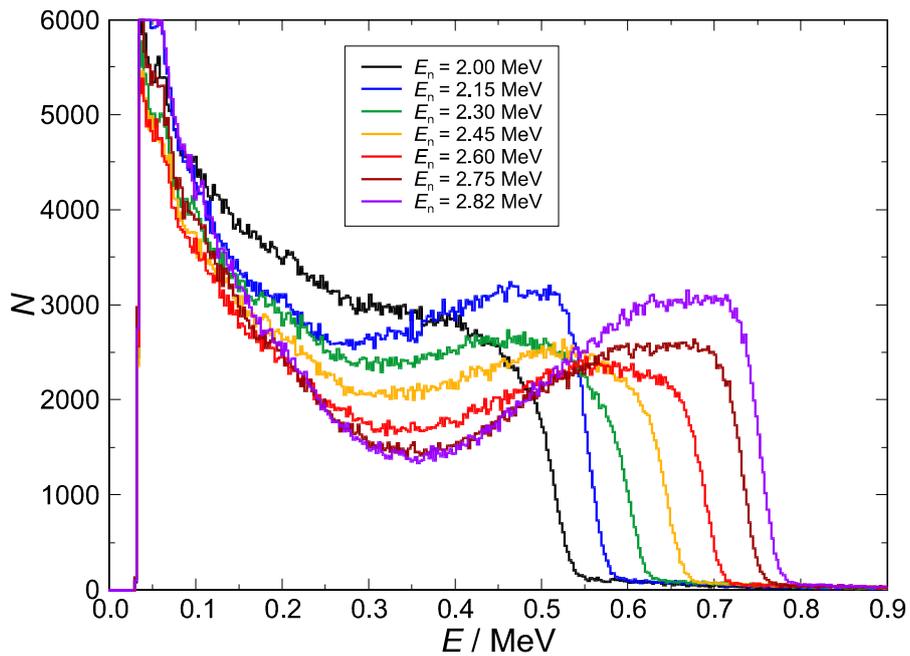

Figure 8: Number of counts, *N*, as a function of the energy, *E*, deposited in the detector for neutrons with energies from 2.0 MeV to 2.8 MeV produced by the T(p,n)$^3$He-reaction with proton beams from 2.8 MeV to 3.6 MeV. All measurements are normalized to the same number of counts from a monitor detector.